\documentclass[twocolumn,showpacs,preprintnumbers,amsmath,amssymb]{revtex4}

\usepackage{graphicx}
\usepackage{dcolumn}
\usepackage{bm}

\begin{document}

\preprint{To be published in Physical Review Letters}

\title{Current-induced two-level fluctuations in pseudo spin-valves
(Co/Cu/Co) nanostructures}

\author{A. F\'{a}bi\'{a}n, C. Terrier, S. Serrano Guisan, X. Hoffer, M. Dubey,
L. Gravier, and J.-Ph. Ansermet} \affiliation{Institut de Physique des
Nanostructures, \'{E}cole Polytechnique F\'{e}d\'{e}rale de Lausanne, CH-1015
Lausanne, Switzerland}
\author{J.-E. Wegrowe}
\affiliation{\'{E}cole Polytechnique, Laboratoire des Solides Irradi\'{e}s,
F-91128 Palaiseau Cedex, France}

\date{October 9, 2003}

\begin{abstract}
Two-level fluctuations of the magnetization state of pseudo spin-valve pillars
Co(10~nm)/Cu(10~nm)/Co(30~nm) embedded in electrodeposited nanowires ($\sim$
40~nm in diameter, 6000~nm in length) are triggered by spin-polarized currents
of $10^{7}$~A/cm$^{2}$ at room temperature. The statistical properties of the
residence times in the parallel and antiparallel magnetization states reveal
two effects with qualitatively different dependences on current intensity. The
current appears to have the effect of a field determined as the bias field
required to equalize these times. The bias field changes sign when the current
polarity is reversed. At this field, the effect of a current density of 10$^7$
A/cm$^2$ is to lower the mean time for switching down to the microsecond range.
This effect is independent of the sign of the current and is interpreted in
terms of an effective temperature for the magnetization.
\end{abstract}

\pacs{75.40.Gb, 75.60.Jk, 75.60.Lr}

\maketitle


Current-induced magnetization switching (CIMS) was predicted by
Slonczewski~\cite{slo96} after a first publication by Berger~\cite{ber96}.
Observation of this phenomenon in several sample configurations was reported a
few years later : homogeneous Ni nanowires~\cite{weg99}, manganite trilayer
junctions~\cite{sun99} and (Co/Cu/Co) sandwich
structures~\cite{mye99,tso98,gro01,sun02}. The potential application of the
latter structure as a non-volatile magnetic memory motivates the development of
detailed models for the theoretical mechanisms underlying CIMS. Most of the
present models~\cite{wai00,hei01,weg00,gro03,zha02} agree on the fact that the
Landau-Lifshitz-Gilbert (LLG) equation can be modified by a current-dependent
term. This term acts either as a torque, an effective field or leads to spin
transfer by a relaxation process.  Two experimental approaches are preferred,
sweeping the magnetic field $H$ or the applied current $I$ in order to obtain
$R(I)$, $R(H)$, $dV/dI(H)$ or $dV/dI(I)$. Alternatively, observations of the
relaxation of the magnetization~\cite{weg02a,alb02,mye02} provide information
on the magnetic energy profile.

Recent experimental work showed that it is possible to produce Two Level
Fluctuation (TLF) in spin-valve nanostructures with the injection of a
spin-polarized current~\cite{mye02,ura03}. In this paper, the TLF produced by
the current are studied in pseudo spin-valves and analyzed in terms of a
potential profile composed of two wells separated by a barrier. The applied
field can be adjusted so that the potential well is symmetrical. At this field
it becomes especially clear that the current enhances the jump rate
irrespective of the current sense.


This study focuses on the irreversible part of the hysteresis in a (Co/Cu/Co)
pseudo spin-valve burried in the middle of a long Cu nanowire. A uniaxial
magnetocrystalline anisotropy can be obtained~\cite{weg02b}. Thus, the
spin-valve behaves as a two-state system defined by the two relative
orientations of the magnetic layers. The samples were produced by the method of
electrodeposition in track-etched membrane templates~\cite{fer99}. Gold layers
were sputtered on both sides of a porous polycarbonate membrane, the pores left
open were filled electrochemically with Co and Cu. Wires of
Cu(1000)/[Co(10)/Cu(10)/Co(30)]/Cu(4950), about 40~nm in diameter and 6000~nm
in length, were obtained. A contact to a single nanowire was established by
monitoring the potential between both sides of the membrane during the
electrodeposition~\cite{weg98}.

\begin{figure}[hbt]
\includegraphics[width=7cm]{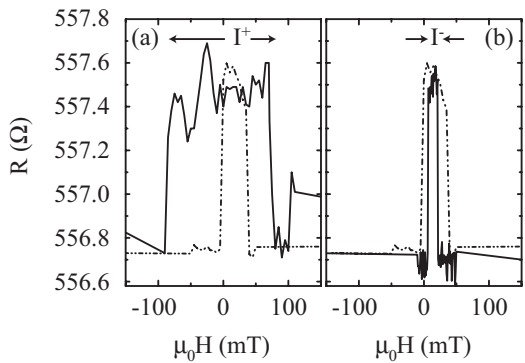}
\caption{\label{fig:cycle} (a) Hysteresis half-cycle at large positive current
(full line) and low current (dashed line). (b) Hysteresis half-cycle at large
negative current (full line) and low current (dashed line). Positive current:
electrons flow from the thin to the thick layer.}
\end{figure}

Experiments were performed at room temperature. For characterization purposes,
giant magnetoresistance (GMR) of the spin-valve system was measured at low
current ($\sim 10^{4}~$A/cm$^{2}$). The sample shape insured
current-perpendicular-to-the-plane (CPP) geometry. The magnetic field was
applied in the direction parallel to the plane of the Co layers. The GMR
results showed a hysteretic behaviour with abrupt steps between two resistance
values $R^{p}$ and $R^{ap}$ ($\Delta R=R^{ap}-R^{p}$ is typically 1~$\Omega$ or
$\Delta R / R_{wire} =$ 0.18\%) corresponding to the switching of the relative
magnetization orientation of the ferromagnetic layers (dashed lines in
Fig.~\ref{fig:cycle}). The abrupt single transitions between these two
orientations suggest single domain structures.

\begin{figure}[hbt]
\includegraphics[width=7cm]{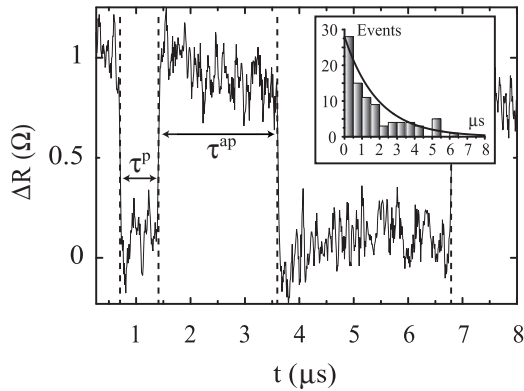}
\caption{\label{fig:trace} Time-resolved response of an applied current of 8
$\mu$s duration. Inset: typical histogram of the residence times $\tau_{ap}$ or
$\tau_{p}$.}
\end{figure}

Hysteresis measurements were also performed under large DC currents ($\sim
10^{7}$~A/cm$^{2}$). Large currents affect the field range over which the spin
valve is in the antiparallel state. This range increases with increasing
positive currents $I^{+}$ (full line in Fig.~\ref{fig:cycle}(a)) and decreases
with negative currents $I^{-}$ (full line in Fig.~\ref{fig:cycle}(b)). The
positive current is defined as the one for which electrons flow from the thin
to the thick magnetic layer (in contrast to the definition of
Ref.~\cite{gro01}). For each current $I^{+}$ or $I^{-}$, we determined the
magnetic field $H_{sw}^{p \rightarrow ap}$ at which a parallel to antiparallel
transition occurred. The time-domain measurements were carried out as follows.
A saturation field of 1~T was established and swept down with a rate of 0.05
T/s to the measurement field $H$ in the vicinity of $H_{sw}^{p \rightarrow
ap}$. At this field value $H$, a current pulse was applied, of amplitude $I$
and of 8 $\mu$s duration. The resistance was recorded as a function of time.
For a broad range of applied fields, the spin valve system presented TLF
between $R^{p}$ and $R^{ap}$ (Fig.~\ref{fig:trace}). The apparent intermediate
steps in Fig.~\ref{fig:cycle} as well as the fluctuations that appears in
Fig.~\ref{fig:trace} are nothing but noise. The stochastic nature of this
process was assessed by determining the histograms of the residence times
$\tau_{ap}$ or $\tau_{p}$ before each transition. They presented an exponential
distribution (inset of Fig.~\ref{fig:trace}) from which the characteristic
times $\langle \tau_{ap} \rangle$ and $\langle \tau_{p} \rangle$ could be
extracted.

\begin{figure}[hbt]
\includegraphics[width=7cm]{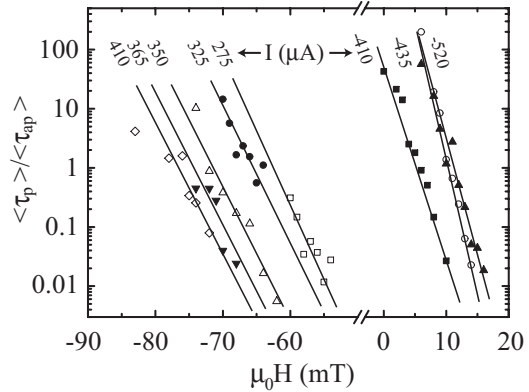}
\caption{\label{fig:tau_up_tau_down} Ratio $\langle \tau_{p} \rangle / \langle
\tau_{ap} \rangle$ vs. magnetic field for different applied currents.}
\end{figure}

For each current value, the protocol was repeated for several values of $H$
until $H$ was so far from $H_{sw}^{p \rightarrow ap}$, that the fluctuations
were too scarce. This measurement process was repeated for several current
values (either positive or negative). This protocol and the measurement setup
implies two limiting currents: the lowest current is the one that allows the
observation of at least one magnetization switching (p $\rightarrow$ ap
$\rightarrow$ p) in the interval time of 8 $\mu$s, and the highest current is
the one for which the switching rate is below the bandwidth of the measurement
setup, 25~MHz. We collected data of average times versus field and current
$\langle \tau_{p} \rangle / \langle\tau_{ap} \rangle$
(Fig.~\ref{fig:tau_up_tau_down}) at fields around the p $\rightarrow$ ap
transition seen in the high current GMR curves (Fig.~\ref{fig:cycle}). We have
carried out a detailed study of the TLF around the transition $H_{sw}^{p
\rightarrow ap}$.

We can assume that the thick Co layer remains fixed because we restricted the
magnetic field values to a small enough range. The metastable characteristics
of the magnetization of the thin Co layer can be described by the
N\'{e}el-Brown activation process~\cite{coffey}. The energy barrier depends on
the shape and magnetocrystalline anisotropy, on the external field and on the
dipole field due to the pinned layer~\cite{weg02b}. At low current, the
spin-valve is stable and does not show TLF. The effective barrier when the
double well is symmetric can be estimated to be of the order of several
thousands Kelvin~\cite{wer95}. Therefore the TLF at large current cannot be
ascribed to Joule heating and must arise from spin polarization of the current.

\begin{figure}[hbt]
\includegraphics[width=7cm]{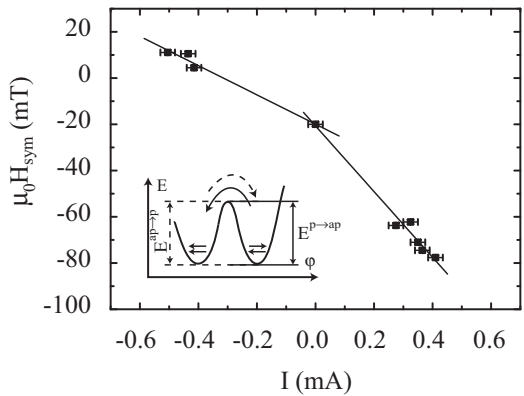}
\caption{\label{fig:Hsym} External magnetic field $H_{sym}(I)$ vs. applied
current $I$ at which $\langle \tau_{ap}\rangle = \langle \tau_{p} \rangle$.
Inset: schematic view of the potential profile, where $\varphi$ represents the
relative magnetization orientation of the Co layers. Straight lines are guide
to the eye.}
\end{figure}

The mean time to escape from a local energy minimum $i$ over an effective
barrier into another local minimum $j$, where $\{ i,j \} =\{ p,ap \}$ or $\{
ap,p \}$ (inset in Fig.~\ref{fig:Hsym}), can be written in the form of an
N\'{e}el-Brown law~\cite{cof98}
\begin{equation}
\tau _{i}=\tau _{0} \exp{\left( \frac{E^{i \rightarrow j}(H)}{k_{B}T} \right) }
\end{equation}
\noindent where $E^{i \rightarrow j}$ is the energy maximum of the barrier
measured from the local minimum $i$ and $\tau_0$ the waiting time at zero
energy barrier. The value of $\tau_0$ we choose is not critical for the outcome
of our analysis. We set it at $\tau_0 \sim$ 0.1~ns as a reasonable order of
magnitude.

We report in Fig.~\ref{fig:Hsym} the value of the magnetic field $H_{sym}(I)$
applied at each current $I$ in order to obtain a symmetric profile, that is,
when $\langle \tau_{ap}\rangle / \langle \tau_{p} \rangle =1$. The bias field
necessary to make the magnetic potential well symmetric is a monotonic, almost
linear function of the current. Here, the effect of a positive current appears
as a positive bias field, since a negative field must be applied to compensate
for it so as to keep the well symmetric. A positive bias field corresponds to a
tendency to remain in the antiparallel state. This is equivalent to the
hysteresis widening of Fig.~\ref{fig:cycle}.

\begin{figure}[hbt]
\includegraphics[width=7cm]{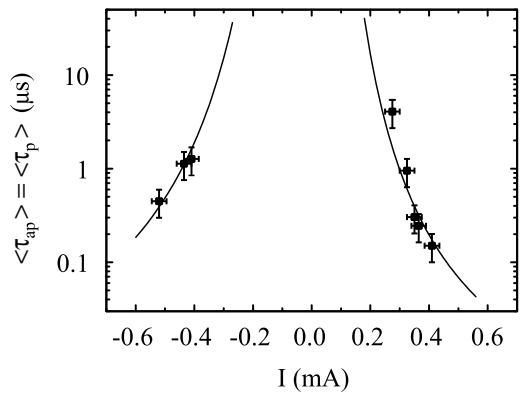}
\caption{\label{fig:energie} Mean time $\langle \tau_{ap}\rangle = \langle
\tau_{p} \rangle$ as a function of the current $I$ at the field $H_{sym}(I)$.
Line: prediction of Eqs.~\ref{eq:rate}-\ref{eq:temp} with parameters as
indicated in the text.}
\end{figure}

Our data show that the effect of the current is not simply a biasing of the
potential profile. We find that, at the field $H_{sym}(I)$, the mean time
$\langle \tau_{ap}\rangle = \langle \tau_{p} \rangle$ decreases with the
absolute value of the current as shown in Fig.~\ref{fig:energie}. Hence, this
effect is qualitatively different than the dependence $H_{sym}(I)$.

We discuss now the possible interpretations that may account for our
observations of $H_{sym}(I)$ and $\langle \tau_{ap} \rangle (I)$ when $\langle
\tau_{ap} \rangle = \langle \tau_{p} \rangle$. Assuming the injection of
spin-polarized current~\cite{slo96}, the LLG equation can be written
as~\cite{zha03}
\begin{eqnarray}
\frac{d \boldsymbol{M}}{dt} = && - \gamma \boldsymbol{M} \times
\boldsymbol{H}_{eff} + \frac{ \alpha }{M_s} \boldsymbol{M} \times
\frac{d \boldsymbol{M}}{dt} \nonumber\\
&&+ \frac{\gamma a_J}{M_s} \boldsymbol{M} \times \left( \boldsymbol{M} \times
\hat{\boldsymbol{M}}_p \right)\,,
\end{eqnarray}
\noindent where $\boldsymbol{M}$ is the magnetization of the free layer,
$\gamma$ is the gyromagnetic factor, $\boldsymbol{H}_{eff}$ the effective
field, including the applied field, anisotropy field, demagnetizing field and
random fields (caused by thermal fluctuations), $\alpha$ the Gilbert damping
factor, $M_s$ the magnetization value at saturation, $a_J$ the dependence of
the current-driven torque on the applied current, and $\hat{\boldsymbol{M}}_p$
a unit vector representing the magnetization orientation of the pinned layer.
It has been shown that the N\'{e}el-Brown's relaxation formula can be applied
by introducing an activation energy defined as the difference between the true
energy barrier and the work done by the current-driven torque~\cite{zha03}.
Depending on the current, this work can be either positive or negative. This
point of view fits with our observation of $H_{sym}(I)$ which is positive or
negative depending on the polarity. However this model cannot account for the
positive slope at negative current (Fig.~\ref{fig:energie}). Therefore we need
to turn to another mechanism to explain $\langle \tau_{ap} \rangle (I)$ when
$\langle \tau_{ap} \rangle = \langle \tau_{p} \rangle$.

Several authors have considered the excitation of spin waves by
current~\cite{tso02,rez00}. Here, we estimate the effect of the excitation of
spin waves caused by the injection of spin polarized currents in terms of an
effective magnetization temperature, an idea simultaneously raised by Urazhdin
et al.~\cite{ura03}. Electrons are injected in the thin layer with a
polarization $\beta$. Spins of electrons with $s$ character are rapidly relaxed
via spin-orbit scattering, while spins of $d$ electrons relax as the
magnetization. Each $d$ electron carries one Bohr magneton whose relaxation
produces, on average, the equivalent of one magnon. Hence the time rate of
generation of magnons by a current $I$ is counted to be $\alpha_{d} \beta
(I/e)$, where $\alpha_{d}$ is thought of as a coefficient between 0 and 1 that
represents the proportion of  $d$-type conduction electrons. In the order of
magnitude estimate below, we take $\alpha_{d} = 4$\%, $\beta(I>0)$ = 40\% and
$\beta(I<0)$ = 27\%. The dependence of $\beta$ on the sense of the current can
be expected since the spin-valve is asymmetric~\cite{zha02}.

Magnetic resonance studies of the bottleneck effect~\cite{blo50,kit56} tell us
that $d$ electrons relax on a time scale $\tau_{d}$ of the order of 1~ns. So
the average number of magnons follows a rate equation
\begin{equation}
\frac{dn}{dt} = - \frac{n}{\tau_{d}}.
\end{equation}

\noindent The current pulse is very long compared to magnetization dynamics.
Since we detect TLF over a time scale of microseconds, the magnetic excitations
have reached a stationary state out of equilibrium. The average number of
magnons is the one that balances the generation of magnons by the current and
their relaxation to the lattice
\begin{equation}
\frac{n}{\tau_{d}} = \alpha_{d} \beta \frac{I}{e}\,. \label{eq:rate}
\end{equation}

\noindent The average number of spin-waves at a temperature $T_m$ follows a
Bose-Einstein distribution $\left[ \exp ( \hbar \omega / k_{B} T_m ) - 1
\right] ^{-1}$. Taking the spin wave dispersion relation $\hbar \omega (k)
\thickapprox 2JSa^2 \cdot k^2$, where $\boldsymbol{k}$ is the wave vector and
$a$ the lattice constant, the density of magnons at this temperature can be
estimated from~\cite{aharoni}
\begin{equation}
\frac{1}{V} \sum_{k}\langle n_{k} \rangle = \frac{1}{(2 \pi )^{2}}
\left(\frac{k_{B}T_m}{2JSa^{2}} \right)^{3/2} \frac{1}{2}\sqrt{\pi} ~ \zeta
\left( 3/2 \right)\, ,  \label{eq:magnon}
\end{equation}

\noindent where the stiffness constant $2JSa^2$ is of the order of
5~meV$\cdot$nm$^2$ for Co, and the Zeta function $\zeta \left( 3/2 \right)=
2.61$. We can account for the data of Fig.~\ref{fig:energie} by assuming that
this random, incoherent generation of magnons gives rise to a magnetic state of
excitation characterized by a temperature $T_m$ calculated with
Eq.~\ref{eq:rate} and \ref{eq:magnon}. The mean times are assumed to follow

\begin{equation}
\langle \tau _{ap} \rangle (I)= \langle \tau _{p} \rangle (I)=\tau _{0} \exp{
\left( \frac{E_0}{k_{B}T_m(I)} \right) }\, , \label{eq:temp}
\end{equation}

\noindent with $E_0 = 6'300$~K, a reasonable value for a Co layer of this
size~\cite{wer95}. Thus we can account for our data (Fig.~\ref{fig:energie})
with a variation of the effective temperature $T_m(I)$ from about 500~K to
1100~K.


In conclusion, we have measured the current dependence of the magnetic energy
profile of a (Co/Cu/Co) nanopillar. Positive current shifts the field range
over which the TLF zone is seen to more positive magnetic fields, while
negative current shift it to more negative fields. However, both current
directions decrease the jump rate. Consequently, these two qualitatively
different features cannot be accounted for with a current dependent effective
torque only. Instead, it appears that an irreversible transfer of magnetic
momentum occurs, leading to spin-wave excitations.

We acknowledge support from the Swiss NSF through grant 200020-100271, from EC
grant EuNITT PPRN-CT-2000-00047 through OFES 99.0141 and EC grant NEXT
IST-2001-37334 through OFES 02.0265.

\bibliography{fabian}

\end{document}